\documentclass{article}
\usepackage[%
  colorlinks=true,
  urlcolor=blue,
  linkcolor=blue,
  citecolor=blue
]{hyperref}

\makeatletter
\let\cat@comma@active\@empty
\makeatother

\usepackage{mathtools,amsmath,amssymb,amsfonts,dsfont,mathrsfs}
\usepackage{centernot}
\usepackage{xcolor}
\hypersetup{linktocpage,colorlinks=true,urlcolor=blue!80!red,linkcolor=blue,citecolor=red}
\usepackage{siunitx}
\usepackage{array}
\usepackage{braket}

\usepackage{etoolbox}
\usepackage{breqn}
\usepackage{authblk}

\newcommand{\Ga}{\Gamma}

\newcommand{\comm}[2]{\left[#1,#2\right]}

\newcommand*{\diag}{\operatorname{diag}}

\newcommand{\dn}[2]{\,{\mathrm{d}}^{#1}\!{#2}\;}

\newcommand\UTFSM{Departamento de F\'\i sica, Universidad T\'{e}cnica Federico Santa Mar\'\i a, \\ Casilla 110-V, Valpara\'\i so, Chile}
\newcommand\CCTVal{Centro Cient\'\i fico Tecnol\'ogico de Valpara\'\i so, \\ Casilla 110-V, Valpara\'\i so, Chile}
\newcommand\CFF{Centro de F\'\i sica Fundamental,  Universidad de los Andes,\\ 5101 M\'erida, Venezuela}

\hypersetup{%
  pdftitle={A polynomial model of purely affine gravity},
  pdfauthor={Aureliano Skirzewski,}{Oscar Castillo-Felisola},
  pdfkeywords={Affine Gravity,} {Torsion,} {Generalised Gravity.},
  pdflang={English}
}

\title{A polynomial model of purely affine gravity}

\author[1,2]{Oscar {Castillo-Felisola}\thanks{o.castillo.felisola@gmail.com}}
\author[3]{Aureliano {Skirzewski}\thanks{askirz@gmail.com}}
\affil[1]{\CCTVal}
\affil[2]{\UTFSM.}
\affil[3]{\CFF.}

\begin{document}

\maketitle

\begin{abstract}
  We present a purely affine gravitational model in four dimensions built up entirely on the bases of full diffeomorphism invariance, and power-counting renormalizability. We show that its non-relativistic limit around a homogeneous and isotropic spacetime yields to a Newtonian gravity.

  \noindent\texttt{PACS No.:} 04.25.Nx, 04.50.Kd, 04.90.+e\\
  \noindent\texttt{Keywords:} Affine Gravity, Torsion, Generalised Gravity.
\end{abstract}

\section{Introduction}

In a critique to Newtonian mechanics, Mach proposed that inertial forces should have a dynamical rather than a kinematical origin (for a deeper discussion on the subject of Mach's principle see Ref.~\cite{Lichtenegger:2004re} and references therein).

Notice that any locally Minkowskian metric in the kinematics of the description of spacetime will introduce a notion of inertial forces at a microscopic level~\cite{Sciama:1964wt}. With this in mind, we will explore the dynamical origin of inertial forces, studying the dynamics of the affine connection of a manifold with torsion. For this end, we use the most general power-counting renormalizable action that includes only the gauge connection associated with diffeomorphisms invariance.

During the last years an increasing amount of alternative theories of gravity have been built and tested. Yet, General Relativity (GR) has proven to be the most successful theory of gravity.  Still,  it is not as successful as we may wish~\cite{DeFelice:2010aj,Capozziello:2011et,Kiefer:2013jqa}. Part of the problem is that the standard quantization procedure cannot be applied  properly on GR. Moreover, not only it is not renormalizable, but there are  problems with the choice of variables to be quantized and the choice of the Hilbert space to be used. Although we dare not to say anything against metric spacetimes, to sum over all possible field configurations of the metric seems to be wrong, as this would imply summing Euclidean and Minkowski like contributions to the transition amplitudes on equal terms. Additionally, we might also consider the difficulties of  quantizing  non-polynomial field theories, and more specifically square roots of the metric that appears in the Hamiltonian in an ADM formulation of GR.

In order to bypass some of these issues, several approaches have been designed that use the connection as a fundamental field. For instance, a well-known example comes from the context of Cartan formulations of gravity,\cite{Olmo:2011uz} using the relation between the Weitzenb\"ock and Levi-Civita connections it is possible to obtain an equivalent Lagrangian to the one by Einstein and Hilbert, as a function of the torsion field. This approach is known as Teleparallel Gravity~(see Ref.~\cite{Maluf:2013gaa,Teleparallel,Baez:2012bn} and references within).

Furthermore, another alternative description of GR developed initially by Ashtekar uses the spin connection as the fundamental field and the frame field turns out to be its canonically conjugated momentum. In the context of Loop Quantum Gravity (LQG), using Ashtekar connection, a successful quantization program has been achieved~\cite{Ashtekar:2004eh,thiemann2007loop}. Originally, this approach towards quantum gravity addressed the concerns of the quantization of non-polynomial functions of the gravitational field, but later on it turned out that diffeomorphisms symmetry would not show up when the quantum operators were not of the correct density weight, which forces one to reintroduce the squared root~\cite{Thiemann:1996aw}.  Some  strength of this quantization program lie within a theorem by H. Sahlmann \emph{et al.} in Ref.~\cite{Lewandowski:2005jk} that states the only diffeomorphisms invariant Hilbert space that supports the Heisenberg algebra, for the connection and its associated momentum, is the one of LQG. In spite of its success, LQG has not advanced enough to conclude that its low energy effective description is GR. Currently, there is no clue about the LQG effective description at other scales,  nor its continuum spacetime limit either. Therefore, we cannot conclude that the search for a fundamental theory of gravitational interactions has ended. On the contrary, there are increasingly many alternatives to the usual metric description of gravity and they all must be tested against experiments and observations.\cite{Berti:2015itd}

In this article we study a power-counting renormalizable,  diffeomorphism invariant model  consisting  solely of an affine connection  (with torsion). We expect  this model may  overcome  the uniqueness theorem about diffeomorphism invariant theories of connections, since we  have no fundamental metric field to quantize. The earliest model that argues  a description of gravitational interaction in terms of connections as fundamental fields  was presented by Eddington~\cite{Eddington1923math}, for an spacetime with positive cosmological constant. He proposed the square root of the determinant of the Ricci tensor as the gravitational Lagrangian.

It has also being emphasized the character of GR as a gauge theory in order to address the issues of quantization and regularization, as in LQG.   Authors like N. Pop{\l}awski~\cite{Poplawski:2012bw} and K. Krasnov~\cite{Krasnov:2011pp} have advanced the road towards a pure connection gravity theory.

The article is organized as follows: In Sec.~\ref{sec:3} we analyse the more general ``gravitational'' theory built with the affine connection and power-counting renormalizable. In Sec.~\ref{sec:4} we study the four-dimensional model, built under the same precepts than before. Additionally, we found solutions to the equations of motion assuming a static, homogeneous and isotropic background, and show that in the non-relativistic limit of the theory the gravitational potential is Newtonian. Finally, in Sec.~\ref{sec:dis} we briefly discuss the reaching consequences of the model.

%--------- Scattered Ideas
\section{\label{sec:3} Warming up: The three-dimensional case}

Formally, the curvature of a manifold is defined through the commutator of covariant derivatives under diffeomorphims, $\hat{\nabla}_\mu$, but for general choice of the connection, $\hat{\Ga}^\mu{}_{\nu\lambda}$, there is an extra contribution given by its antisymmetric part in the lower indices, $T^\mu{}_{\nu\lambda} = 2\hat{\Ga}^\mu{}_{[\nu\lambda]}$. Therefore, the commutator of the covariant derivatives acting on a vector, $V^\rho$, yields,
\begin{equation}
  \comm{\hat{\nabla}_{\mu}}{\hat{\nabla}_{\nu}}V^\rho = \hat{R}_{\mu\nu}{}^\rho{}_\lambda V^\lambda - T^\rho{}_{\mu\nu}\nabla_{\rho}V^\rho.
  \label{curvdef}
\end{equation}
Note that $T^\rho{}_{\mu\nu}$ is a nine-dimensional tensor representation under diffeomorphisms.

In order to build topological invariants of density one, we can use the skew-symmetric Levi-Civita tensor $\epsilon^{\mu_1\mu_2\dots\mu_n}$ in $n$-dimensional space(-time).

%% \begin{widetext}
With these ingredients, in a three-dimensional space we write an action 
\begin{dmath}
  \label{accion3d}
  S[\Gamma] =
  \int \dn{3}{x}  \Bigg\{
  \hat{R}_{\mu_1\mu_2}{}^\rho{}_{\mu_3} T^\sigma{}_{\mu_4\mu_5} \sum_{\pi \in  \mathrm{Z}_5}C_\pi\delta_\rho^{\mu_{\pi(1)}} \delta_\sigma^{\mu_{\pi(2)}} \epsilon^{\mu_{\pi(3)}\mu_{\pi(4)}\mu_{\pi(5)}}
  + T^\rho{}_{\mu_1\mu_2} T^\sigma{}_{\mu_3\mu_4} T^\tau{}_{\mu_5\mu_6} \sum_{\pi \in \mathrm{Z}_6}D_\pi\delta_\rho^{\mu_{\pi(1)}} \delta_\sigma^{\mu_{\pi(2)}}\delta_\tau^{\mu_{\pi(3)}}\epsilon^{\mu_{\pi(4)}\mu_{\pi(5)}\mu_{\pi(6)}}
  + T^\rho{}_{\mu_1\mu_2} \hat{\nabla}_{\mu_3} T^\sigma{}_{\mu_4\mu_5}\sum_{\pi \in \mathrm Z_5}E_\pi\delta_\rho^{\mu_{\pi(1)}} \delta_\sigma^{\mu_{\pi(2)}}\epsilon^{\mu_{\pi(3)}\mu_{\pi(4)}\mu_{\pi(5)}} \Bigg\}, 
  %\end{split}
\end{dmath}
where all possible permutations of $n$ elements $\pi \in \mathrm{Z}_n$ have been included in the sums with  different constants $C_\pi$, $D_\pi$ and $E_\pi$ for  permutation. 
%% \end{widetext}

The torsion field can be decomposed into invariant tensors respecting the symmetry,
\begin{equation}
  T^\sigma{}_{\mu\nu} = \epsilon_{\mu\nu\rho} T^{\sigma\rho} + A_{[\mu}\delta^\sigma{}_{\nu]},
\end{equation}
with a symmetric $T^{\sigma\rho}$ of density weight  $w = 1$, and \mbox{$A_\mu = T^\nu{}_{\mu\nu}$} is the trace part of the more arbitrary $T^\sigma{}_{\mu\nu}$.
In the action in Eq.~\eqref{accion3d} an extra Chern--Simons term, which in three dimensions is invariant under diffeomorphisms, with a coefficient $B_9$.

The affine connection can be decomposed into its symmetric and antisymmetric parts, 
\begin{equation}
  \hat{\Gamma}^\lambda{}_{\mu\rho} = {\Gamma}^\lambda{}_{(\mu\rho)} + \epsilon_{\mu\rho\sigma}T^{\lambda\sigma} + A_{[\mu}\delta^\lambda{}_{\rho]},
\end{equation}
where  $\epsilon_{\mu\rho\sigma}$ has been introduced, and it is related to the skew symmetric $\epsilon^{\mu\rho\sigma}$ through the identity \mbox{$\epsilon^{\lambda\mu\nu}\epsilon_{\rho\sigma\tau}=3!\delta^{\lambda}{}_{[\rho}\delta^\mu{}_{\sigma}\delta^{\nu}{}_{\tau]}$.} Therefore, the curvature tensor can be expressed as 
\begin{dmath}
  \label{RiemmanDecomposition}
  \hat{R}_{\mu\nu}{}^\sigma{}_\rho=
  {R}_{\mu\nu}{}^\sigma{}_\rho
  -2\epsilon_{\rho\alpha[\mu}\nabla_{\nu]}T^{\sigma\alpha}
  +\partial_{[\mu}A_{\nu]}\delta^\sigma_\rho
  +\delta^\sigma_{[\mu}\nabla_{\nu]}A_\rho
  +\epsilon_{\mu\nu\kappa}T^{\kappa\sigma}A_\rho
  -\delta^\sigma_{[\mu}\epsilon_{\nu]\rho\alpha}T^{\alpha\beta}A_\beta 
  +\frac{1}{2}\delta^\sigma_{[\mu}A_{\nu]}A_\rho
  -2\epsilon_{\alpha\beta[\mu}\epsilon_{\nu]\rho\delta}T^{\sigma\alpha}T^{\beta\delta},
\end{dmath}
where $\nabla_\rho$ and ${R}_{\mu\nu}{}^\lambda{}_\rho$ are the covariant derivative and  curvature associated to the symmetric part of the connection. Notice that Bianchi identity, obtained as $\epsilon^{\mu\nu\lambda} R_{\mu\nu}{}^\rho{}_\lambda=0$, leads us to the following
\begin{equation}
  \label{bianchi}
  \epsilon^{\mu\nu\rho} \hat{R}_{\mu\nu}{}^\lambda{}_\rho = 4\nabla_\rho T^{\rho\lambda}
  +2\epsilon^{\mu\nu\lambda}\partial_\mu A_\nu-4T^{\lambda\rho}A_\rho. 
\end{equation}
Using the  Eqs.~\eqref{RiemmanDecomposition} and~\eqref{bianchi} one can rewrite the action (up to a boundary term) as
\begin{dmath}
  S[\Gamma,T,A] =
  \int \dn{3}{x} \bigg( 
  B_1{R}_{\mu\nu}{}^{\mu}{}_\rho T^{\nu\rho} 
  +B_2\epsilon^{\mu\nu\rho}{R}_{\mu\nu}{}^{\sigma}{}_\sigma A_\rho
  +B_3\epsilon^{\mu\nu\rho}A_\mu\partial_\nu A_\rho
  +B_4T^{\mu\nu}{\nabla}_\mu A_\nu
  +B_5T^{\mu\nu}A_\mu A_\nu
  +B_6\det(T^{\mu\nu}) 
  +B_7\epsilon^{\mu\nu\lambda}\Big({\Gamma}^{\sigma}{}_{\mu\rho}\partial_\nu{\Gamma}^{\rho}{}_{\lambda\sigma}
  +\frac{2}{3}{\Gamma}^{\tau}{}_{\mu\rho}{\Gamma}^{\rho}{}_{\nu\sigma}{}{\Gamma}^{\sigma}{}_{\lambda\tau}{}\Big)
  + B_8\epsilon^{\mu\nu\rho}{\Gamma}^{\sigma}{}_{\mu\sigma}\partial_\nu{\Gamma}^{\tau}{}_{\rho\tau}
  \bigg),
  \label{accion3dfinal}
\end{dmath}
with $B_i$ the coupling constants.

At this point, it is useful to introduce what we have called the ``Eddington's trick''~\cite{Eddington1923math}. First of all, notice that in the usual Einstein--Hilbert action the variation of the action with respect to the Ricci tensor yields an inverse metric density. Thus, a sort of dual theory could be obtained by identifying the tensor density obtained from the variation of the action with respect to the symmetric part of Ricci tensor with the inverse metric density (see Ref.~\cite{Eddington1923math,Poplawski:2012bw})
\begin{equation}\label{metric}
  \frac{\delta\ }{\delta R_{(\mu\nu)}} S[\Gamma] \Longrightarrow \sqrt{g} g^{\mu\nu} .
\end{equation}

Noticing that in  the first term, the variation respect to the Ricci tensor yields to $T^{\mu\nu}$, it can be argued that in a standard theory of gravity this tensor density corresponds to $\sqrt{g}g^{\mu\nu}$.
Therefore, Eq.~\eqref{accion3dfinal} reveals a one to one correspondence with general relativity nonminimally coupled to the $A_\mu$ field,
\begin{dmath}
  \label{accion3dGR}
  S[g,{\Gamma},A] = \int \dn{3}{x} \bigg(
  \sqrt{g} \Big(  B_1 {R} + B_4{\nabla}^\mu A_\mu + B_5 A_\mu A^\mu + B_6  \Big)
  + B_2\epsilon^{\mu\nu\rho} {R}_{\mu\nu}{}^{\sigma}{}_\sigma A_\rho
  + B_3\epsilon^{\mu\nu\rho}A_\mu\partial_\nu A_\rho
  + B_7\epsilon^{\mu\nu\lambda}\Big({\Gamma}^{\sigma}{}_{\mu\rho}\partial_\nu{\Gamma}^{\rho}{}_{\lambda\sigma}
  + \frac{2}{3}{\Gamma}^{\tau}{}_{\mu\rho}{\Gamma}^{\rho}{}_{\nu\sigma}{}{\Gamma}^{\sigma}{}_{\lambda\tau}{}\Big)
  + B_8\epsilon^{\mu\nu\rho}{\Gamma}^{\sigma}{}_{\mu\sigma}\partial_\nu{\Gamma}^{\tau}{}_{\rho\tau}
  \bigg)
\end{dmath}
Thus, an interesting sector of the theory corresponds to the space of non-degenerated $T^{\mu\nu}$. 

\section{\label{sec:4} Four-dimensional metricless (and torsionful) action}

Following the precepts  already stated, we start  by defining an irreducible representation decomposition for the full connection field 
\begin{equation}
  \hat{\Gamma}^\mu{}_{\rho\sigma} = {\Gamma}^\mu{}_{\rho\sigma} + T^\mu{}_{\rho\sigma} = {\Gamma}^\mu{}_{\rho\sigma} + \epsilon_{\rho\sigma\lambda\kappa}T^{\mu,\lambda\kappa}+A_{[\rho}\delta^\mu_{\nu]},
\end{equation}
where ${\Gamma}^\mu{}_{\rho\sigma}$ denotes a forty-dimensional symmetric connection, $A_\mu$ is a four-dimensional vector field  that gives trace to the antisymmetric part of the full connection, and  $T^{\mu,\lambda\kappa}$ is a twenty-dimensional Curtright field (see Ref.~\cite{Curtright:1980yk}) that is defined through the symmetry of its indices: antisymmetric in the last two indices, and it has a cyclic property $T^{\mu,\lambda\kappa}+T^{\lambda,\kappa\mu}+T^{\kappa,\mu\lambda}=0$. In other words that $T^{[\mu,\lambda]\kappa}=\frac{1}{2}T^{\kappa,\lambda\mu}$, just as for  the Riemmann tensor ${R}_{\mu[\nu}{}^\alpha{}_{\lambda]}=\frac{1}{2}{R}_{\lambda\nu}{}^\alpha{}_{\mu}$. Notice that due to its symmetries, the contraction $\epsilon_{\rho\sigma\lambda\kappa}T^{\mu,\lambda\kappa}$ is traceless.

Additionally, since no metric is present  the epsilon symbols are not related by lowering or raising their indices, but instead one demands that $$\epsilon^{\delta\eta\lambda\kappa}\epsilon_{\mu\nu\rho\sigma}=4!\delta^{\delta}{}_{[\mu}\delta^\eta{}_{\nu}\delta^{\lambda}{}_{\rho} \delta^\kappa{}_{\sigma]}.$$

%%\begin{widetext}
One can  write all the combinations of fields that would presumably be renormalizable with these three independent fields ---up to a boundary term---,
\begin{dmath}[compact, spread=2pt] 
  \label{4dfull}
  S[{\Gamma},T,A] =
  \int\dn{4}{x}\Bigg[
    B_1  R_{\mu\nu}{}^{\mu}{}_{\rho} T^{\nu,\alpha\beta}T^{\rho,\gamma\delta}\epsilon_{\alpha\beta\gamma\delta}
    +B_2 \Big( R_{\mu\nu}{}^{\sigma}{}_\rho+\frac{2}{3}\delta^\sigma{}_{[\mu} R_{\nu]\lambda}{}^{\lambda}{}_\rho \Big) T^{\beta,\mu\nu}T^{\rho,\gamma\delta}\epsilon_{\sigma\beta\gamma\delta}
    +B_3  R_{\mu\nu}{}^{\mu}{}_{\rho} T^{(\nu,\rho)\sigma}A_\sigma
    + B_4\Big( R_{\mu\nu}{}^{\sigma}{}_\rho+\frac{2}{3}\delta^\sigma{}_{[\mu} R_{\nu]\lambda}{}^{\lambda}{}_\rho \Big)\Big(T^{\rho,\mu\nu}A_\sigma-\frac{1}{4}\delta^\rho_\sigma T^{\kappa,\mu\nu}A_\kappa\Big)
    +B_5 R_{\mu\nu}{}^{\rho}{}_\rho T^{\sigma,\mu\nu}A_\sigma
    +C_1  R_{\mu\nu}{}^{\mu}{}_{\rho} \nabla_\sigma T^{(\nu,\rho)\sigma}
    +C_2 R_{\mu\nu}{}^{\rho}{}_\rho \nabla_\sigma T^{\sigma,\mu\nu} 
    +D_1T^{\alpha,\mu\nu}T^{\beta,\rho\sigma}\nabla_\gamma T^{(\lambda, \kappa) \gamma}\epsilon_{\beta\mu\nu\lambda}\epsilon_{\alpha\rho\sigma\kappa}
    +D_2T^{\alpha,\mu\nu}T^{\lambda,\beta\gamma}\nabla_\lambda T^{\delta,\rho\sigma}\epsilon_{\alpha\beta\gamma\delta}\epsilon_{\mu\nu\rho\sigma}
    +D_3T^{\mu,\alpha\beta}T^{\lambda,\nu\gamma}\nabla_\lambda T^{\delta,\rho\sigma}\epsilon_{\alpha\beta\gamma\delta}\epsilon_{\mu\nu\rho\sigma}
    +D_4T^{\lambda,\mu\nu}T^{\kappa,\rho\sigma}\nabla_{(\lambda} A_{\kappa)} \epsilon_{\mu\nu\rho\sigma}
    +D_5T^{\lambda,\mu\nu}\nabla_{[\lambda}T^{\kappa,\rho\sigma} A_{\kappa]} \epsilon_{\mu\nu\rho\sigma}
    +D_6T^{\lambda,\mu\nu}A_\nu\nabla_{(\lambda} A_{\mu)}
    +D_7T^{\lambda,\mu\nu}A_\lambda\nabla_{[\mu} A_{\nu]} 
    +E_1\nabla_{(\rho} T^{\rho,\mu\nu}\nabla_{\sigma)} T^{\sigma,\lambda\kappa}\epsilon_{\mu\nu\lambda\kappa}
    +E_2\nabla_{(\lambda} T^{\lambda,\mu\nu}\nabla_{\mu)} A_\nu
    +T^{\alpha,\beta\gamma}T^{\delta,\eta\kappa}T^{\lambda,\mu\nu}T^{\rho,\sigma\tau}
    \big(\Lambda_1\epsilon_{\beta\gamma\eta\kappa}\epsilon_{\alpha\rho\mu\nu}\epsilon_{\delta\lambda\sigma\tau}
    +\Lambda_2\epsilon_{\beta\lambda\eta\kappa}\epsilon_{\gamma\rho\mu\nu}\epsilon_{\alpha\delta\sigma\tau}\big) 
    +\Lambda_3 T^{\rho,\alpha\beta}T^{\gamma,\mu\nu}T^{\lambda,\sigma\tau}A_\tau \epsilon_{\alpha\beta\gamma\lambda}\epsilon_{\mu\nu\rho\sigma}
    +\Lambda_4T^{\eta,\alpha\beta}T^{\kappa,\gamma\delta}A_\eta A_\kappa\epsilon_{\alpha\beta\gamma\delta}\Bigg],
\end{dmath}
where the

terms $B_2$ and $B_4$  contain a traceless contribution of the curvature.
%% \end{widetext}
In this case, the induced  ``inverse metric density'' [see Eq.~\eqref{metric}] is 
\begin{dmath}
  \label{4dMetric}
  \bar{g}^{\mu\nu} \equiv \sqrt{g}g^{\mu\nu} = B_1 T^{\mu,\lambda\kappa}T^{\nu,\rho\sigma}\epsilon_{\lambda\kappa\rho\sigma} + B_3 T^{(\mu,\nu)\lambda}A_\lambda + C_1 {\nabla}_\lambda T^{(\mu,\nu)\lambda}.
\end{dmath}

\subsection*{Symmetric solution to the equations of motion}

In four dimensions there is no obvious equivalence of Eq.~\eqref{4dfull} with GR, specially due to the lack of a fundamental metric field in the given model. However, both models are explicitly invariant under diffeomorphisms, and even if their structures and number of degrees of freedom differ, the action in Eq.~\eqref{4dfull} provides a context where parallel transport of particle's velocities on a purely torsional background is nontrivial.

Here, we wish to stablish the model's non-relativistic (Newtonian) limit for the ``geodesic" deviation of ``inertial" observers at rest with respect to a static, isotropic, homogeneous and spatially flat background within the context provided by Eq.~\eqref{4dfull}. In order to properly analyse the model, we  propose the following decomposition of the fields
\begin{align}
  A_\mu &= \delta_\mu^0 A + a_\mu,\\
  T^{\mu,\nu\rho} &= \delta^{\mu}_m\delta^{\nu\rho}_{m0}T + t^{\mu,\nu\rho},\\
  \shortintertext{and}
  \Gamma^\lambda{}_{\mu\nu} &= E \delta^\lambda_0 \delta^m_\mu \delta^m_\nu + F \delta^\lambda_m \delta^m_{(\mu}\delta^0_{\nu)} + G\delta^\lambda_0 \delta^0_{\mu}\delta^0_{\nu} + \gamma^\lambda{}_{\mu\nu},
  \label{GammaExp}
\end{align}
where $\delta^{\mu\nu}_{\lambda\kappa}=\delta^{\mu}_{\lambda}\delta^{\nu}_{\kappa}-\delta^{\mu}_{\kappa}\delta^{\nu}_{\lambda}$.

In order to make perturbation theory we will expand around a static, isotropic and homogeneous solution of the equations of motion, because these are characteristic of the observable universe.

The induced metric in Eq.~\eqref{4dMetric}  on the background is
\begin{dmath}
  \label{3+1metric}
  \sqrt{-g}g^{\mu\nu} = \left(B_3 A + \frac{1}{2}C_1 F\right) T \delta^\mu_m \delta^\nu_m - 3 C_1 E T \delta^\mu_0\delta^\nu_0,
\end{dmath}
while the Ricci curvature tensor calculated from Eq.~\eqref{GammaExp} is
\begin{dmath}
  R_{\mu\nu} = \frac{1}{2} E F \delta^m_\mu \delta^m_\nu - \frac{3}{4} F^2 \delta^0_\mu \delta^0_\nu.
\end{dmath}
Therefore, whether the four-dimensional Eddington's metric structure is Riemannian or pseudo-Riemannian will depend exclusively on the values of the parameters of the action in Eq.~\eqref{4dfull} and the signs of the components of the connection field. 
%\begin{widetext}
The first order perturbations of the action yields
\begin{dmath}[compact, spread=2pt]
  \label{EOM0thOrder}
  \delta S =
  \bigg( \Big( ( B_3 + \frac{8}{3}\, B_4 + \frac{1}{2}\, E_2) A + 4\, C_1  F - 2\, C_1  G \Big) E + 8\, ( - D_1 + 2\, D_2 + D_3) T^2 \bigg) T \delta{\Gamma}^{m}\,_{0 m}
  + \bigg( ( \frac{1}{2}\, B_3 + \frac{4}{3}\, B_4 + \frac{1}{4}\, E_2) A F + ( B_3 - \frac{4}{3}\, B_4 - \frac{1}{2}\, E_2) A G + C_1  F^2 - C_1 F G - D_6 A^2 \bigg) T \delta{\Gamma}^{0 m}\,_{m}
  + \bigg( \Big(- (\frac{1}{2}\, B_3 + \frac{4}{3}\, B_4 + \frac{1}{4}\, E_2) A F + ( - B_3+ \frac{4}{3}\, B_4 + \frac{1}{2}\, E_2) A G - C_1  F^2 + C_1 F G + D_6 A^2 \Big) E + \Big( 12\, ( D_1 - 2\, D_2 - D_3) F + 24\, L_3 A \Big) T^2 \bigg)\delta{T}_{m}\,^{0 m}
  + \bigg( ( 3\, B_3 - 4\, B_4 - \frac{3}{2}\, E_2) A - 3\, C_1 F \bigg) E T \delta{\Gamma}^{0}\,_{0 0}
  + \bigg( 3\Big( - 2\, D_6 A + ( \frac{1}{2}\, B_3 + \frac{4}{3}\, B_4 + \frac{1}{4}\, E_2) F + ( B_3 - \frac{4}{3}\, B_4 - \frac{1}{2}\, E_2) G \Big) E - 24\, L_3 T^2 \bigg) T \delta{A}_{0}=0,
\end{dmath}
%% \end{widetext}
and we are most interested in solutions to the connection field whose contribution to the parallel transport equation of a test particle's velocity is that of a free particle, at least at the low velocity regime
\begin{equation} 
  \ddot{x}^i+2F\dot{x}^0\dot{x}^i=0, \quad \text{and} \quad \ddot{x}^0 + E \, (\dot{x}^i)^2 + G \, (\dot{x}^0)^2 = 0,
\end{equation}
which we can achieve by setting $F=G=0$ and $E \neq 0$ since $(\dot{x}^i)^2$ is already second order in the velocities. Thus, looking again at the equations of motion we can find a nontrivial solution if we set all coupling constants to zero but $B_3 \neq 0$, $B_4 = -\tfrac{3}{2} B_3$, $C_1\neq 0$ and $E_2= 6 B_3$. 

Additionally, we can incorporate perturbative inhomogeneous sources to the connection field equations and check on how these fluctuations affect motion. For this, we consider a matter's action, whose dependence on the affine connection can be almost arbitrary. However, we will presume that it will depend only on the barred metric in Eq.~\eqref{4dMetric}
$$ S_{\text{Matter}} = {S}_{\text{Matter}}[\bar{g}^{\mu\nu}].$$
Thus, a non-moving matter point particle at the origin of the reference frame will contribute to the equations of motion for the gravitational field through the component $\bar{g}^{00}$ following the symmetries of the matter source 
\begin{dmath}
  \label{mattervariation}
  \delta {S}_{\text{Matter}} =  C_1 \Big(- \frac{1}{2} ({\delta\Gamma}^{0}{}_{m n})  T {\delta}^{m n} - \frac{1}{2}  ({\delta T}^{0 0 m})  \imath {p}_{m} + \frac{1}{2}  ({\delta T}^{m 0 n})  E {\delta}_{m n} \Big)\frac{\partial\mathcal{L}_{\text{Matter}}}{\partial \bar{g}^{00}}.
\end{dmath}

\subsection*{Scalar modes  and Newtonian limit}

In order to obtain the non-relativistic limit, \emph{i.e.}, the Newtonian potential, one performs the scalar mode perturbative expansion. One proceeds by substituting the connection and torsion components by their scalar perturbation decomposition,
\begin{equation}
  a_\mu \to \delta_\mu^0 a+\delta_\mu^m \partial_{m}a_L,
\end{equation}
\mbox{}
\begin{dmath}
  t^{\mu,\nu\rho} \to \delta^{\mu}_m\delta^{\nu\rho}_{n0} \Big(t \delta^{m n} + \partial^m \partial^n t_L \Big)
  +\delta^{\mu}_0 \delta^{\nu\rho}_{m0} \partial^m c_L
  + \Big(\delta^{\mu}_0\delta^{\nu\rho}_{mn}-\delta^{\mu}_m\delta^{\nu\rho}_{n0}\Big)\epsilon^{m n p} \partial_{p} b
  +\delta^{\mu}_m \delta^{\nu}_{n} \delta^{\rho}_{p} \Big(\epsilon^{n p q}\partial_q \partial^m d_1 +  (\delta^{m n} \partial^p - \delta^{m p} \partial^n)d_2\Big)
\end{dmath}
and

\begin{dmath}
  \gamma^\lambda_{\mu\nu} \to
  \delta^\lambda_0\delta^0_\mu\delta^0_\nu u 
  + \delta^\lambda_m \delta^0_\mu\delta^0_\nu \partial^m v_L
  + 2\delta^\lambda_0 \delta^0_{(\mu}\delta^m_{\nu)} \partial_m w_L
  + \delta^\lambda_0 \delta^m_\mu\delta^n_\nu \Big(x \delta_{mn} + \partial_m \partial_n x_L\Big)
  + 2\delta^\lambda_m \delta^0_{(\mu}\delta^n_{\nu)} \Big(y_1 \delta^m{}_n + \epsilon^{m p}{}_{n} \partial_p y_2 + \partial^m \partial_n y_L\Big)
  + \delta^\lambda_m \delta^n_{\mu}\delta^p_{\nu} \Big(\delta_{n p} \partial^m z_1 + (\delta^m{}_n \partial_p+\delta^m{}_p \partial_n) z_2 +  (\epsilon^{m q}{}_n \partial_p+\epsilon^{m q}{}_p \partial_n) \partial_q z_3 + \partial^m \partial_n \partial_p z_L\Big),
\end{dmath}
where the scalar fields identified with the sub-index ``L'' correspond to longitudinal degrees of freedom. Vector and tensor perturbations are left for further investigations of the structure of the model.

The first order perturbative expansion of the equations of motion around the already described background in momentum space with $p_0=0$ is given by
%% \begin{widetext}
\begin{dmath}
  \delta S =
  \Big( - 2 E d_2 + t - p^2 t_L + T w_L + 3 T z_1 + 2 T z_2 - T p^2 z_L \Big) 6 B_3 p^2 {\,\delta A}_{0} 
  + \bigg( - E p^2 d_2 + \frac{1}{2} p^2 t - \frac{1}{2} p^4 t_L - \frac{1}{2} T p^2 w_L - 6 E T y_1 + 2 E T p^2 y_L + \frac{3}{2} T p^2 z_1 + T p^2 z_2 - \frac{1}{2} T p^2 p^2 z_L \bigg) C_1 {\,\delta\Gamma}^{0}{}_{0 0} 
  + \bigg(6 B_3 a - \frac{1}{2} C_1 u + 2 C_1 E v_L + \frac{1}{2} C_1 y_1 - \frac{3}{2} C_1 p^2 y_L \bigg) T \imath {p}^{m} {\,\delta\Gamma}^{0}{}_{0 m} 
  + C_1 T p^2 v_L {\,\delta}^{m n} {\,\delta\Gamma}^{0}{}_{m n} 
  + \Big( - p^2 c_L + 2 E E d_2 + 4 E t - 2 E p^2 t_L + 2 E T w_L + 3 T x - T p^2 x_L - 10 E T z_2 + 2 E T p^2 z_L \Big) C_1 \imath {p}_{m} {\,\delta\Gamma}^{m}{}_{0 0} 
  + \bigg( E p^2 d_2 +  \frac{1}{2} p^2 t - \frac{1}{2} p^4 t_L - 2 E T u - \frac{1}{2} T p^2 w_L + 8 E T y_1 - 2 E T p^2 y_L + \frac{1}{2} T p^2 z_1 + T p^2 z_2 - \frac{1}{2} T p^4  z_L \bigg) C_1 {\,\delta}^{m}{}_{n} {\,\delta\Gamma}^{n}{}_{0 m} 
  + \Big( - 4 E d_2  - t +  p^2 t_L + 2 T w_L - 2 E T y_L - 2 T z_1 + T p^2 z_L \Big)  C_1 {p}_{n} {p}^{m} {\,\delta\Gamma}^{n}{}_{0 m} 
  - 6  E T y_2 C_1 \imath {p}^{p} {\epsilon}_{n p}{}^{m} {\,\delta\Gamma}^{n}{}_{0 m} 
  + \bigg(6 B_3 T a + C_1 p^2 d_2 + \frac{1}{2} C_1 T u + C_1 E T v_L - \frac{1}{2} C_1 T y_1 + \frac{1}{2} C_1 T p^2 y_L \bigg) \imath {p}_{m} {\,\delta}^{n p} {\,\delta\Gamma}^{m}{}_{n p} 
  + \Big( - 3 E v_L + y_1 - p^2 y_L\Big) C_1 T \imath {p}^{n} {\,\delta}_{m}{}^{p} {\,\delta\Gamma}^{m}{}_{n p} 
  + \Big( - d_2 + T y_L \Big) C_1 \imath {p}_{m} {p}^{n} {p}^{p} {\,\delta\Gamma}^{m}{}_{n p} 
  + v_L C_1 p^2 \imath {p}_{m} {\,\delta T}^{0 0 m} 
  - v_L C_1 E p^2 {\,\delta}_{m n} {\,\delta T}^{m 0 n} 
  + \bigg( - 6  B_3 a - \frac{1}{2}  C_1 u - C_1 E v_L - \frac{1}{2}  C_1 y_1 - \frac{1}{2}  C_1 p^2 y_L \bigg) {p}_{m} {p}_{n} {\,\delta T}^{m 0 n} 
  + \bigg(6 B_3 E a + \frac{1}{2} C_1 E u - C_1 E E v_L + \frac{1}{2} C_1 E y_1 - \frac{3}{2} C_1 E p^2 y_L - C_1 p^2 z_1 \bigg) \imath {p}_{m} {\,\delta}_{n p} {\,\delta T}^{n m p},
\end{dmath}
%% \end{widetext}
which we will add to the variations of the action of the matter from Eq.~\eqref{mattervariation}
and set $\delta S_{\text{total}}=0$.

Solutions to this set of equations are in general a highly difficult problem that concerns twenty equations of motion with twenty scalar fields to be fixed. Yet, knowledge of the value of some of these scalars does not necessarily help to determine how geodesics are affected. For this, we only need $\gamma^i{}_{00}$ and $\gamma^0{}_{00}$ as these provide the first order contributions to the equations
\begin{equation}
  \label{geodesic1stOrder}
  \ddot{x}^i + \gamma^i{}_{00}(\dot{x}^0)^2 = 0,
  \quad \text{and} \quad
  \ddot{x}^0 + \gamma^0{}_{00}(\dot{x}^0)^2 = 0.
\end{equation}
From Eq.~\eqref{geodesic1stOrder} we can conclude that we need only know $\gamma^i{}_{00} = \partial^i v_L$, which we obtain in Fourier space to be

\begin{equation}
  v_L=\frac{1}{2}\frac{\partial\mathcal{L}_{\text{Matter}}}{\partial\bar{g}^{00}}\frac{1}{p^2}.
\end{equation}
In position space, 
\begin{equation}
  v_L = \frac{1}{8\pi} \frac{ \partial\mathcal{L}_{\text{Matter}} }{ \partial \bar{g}^{00} } \frac{1}{|\vec{x}|}
\end{equation}
is the usual Newtonian potential for a massive far off source.

\section{\label{sec:dis} Discussion}

In this paper we have proposed novel model of gravitational interactions with full diffeomorphisms invariance as the main guiding principle, whose fundamental field is an affine connection and no metric field is assumed (nor needed). Surprisingly, in four dimensions, it upholds the correct Newtonian limit, supporting the suspicions that it may describe some aspects of gravitational physics that have not been exposed yet. Still, the model is alien for anyone accustomed to metric spacetimes or their extensions, specially since no local Lorentz structure is present. We also argue that in the absence of fundamental inertial structure, it becomes a natural playground to test the full reaches of Mach's Principle.

Additionally, within the model, all coupling constants turn out to be dimensionless, a property that has been related to scale invariance and conformally invariant theories (see Refs.~\cite{Buchholz:1976hz,Maldacena:2011mk}). In fact, renormalization is intimately related to the scaling properties of a model and it may be worth to study the quantization and renormalizability of this  model. In doing so, we believe the lack of metric may allow the model to bypass the uniqueness of the diffeomorphisms invariant Hilbert space stated in Ref.~\cite{Lewandowski:2005jk}.

Sticking to the classical theory, the analysis of the cosmological implications is needed. In cosmology there are important aspects with unsatisfactory explanations, such as those related to the matter content of the Universe (in particular the dark energy sector), or the large scale structure formation~\footnote{There is a controversy between the experimental results obtained by BICEP2~\cite{Ade:2014xna} and Planck~\cite{Adam:2014bub} in the respect of their interpretation associated with inflation.}. Additionally, other formal aspects of the model remain unknown, such as the proper number of propagating degrees of freedom, and whether or not a duality exists between our model and one of the well-known metric models (say for example Ho\v{r}ava-Lifshitz gravity~\cite{Horava:2009uw,Sotiriou:2010wn}).

We also believe that almost every aspect related to coupling gravity to matter fields can be extrapolated to couplings with the affine connection without making reference to a fundamental metric or a local Lorentz symmetry. This could be  done rewriting the affine connection in terms of a local $GL(4)$ connection
$\omega_{\mu}{}^a{}_b$, relating the two of them through use of a frame field $e_\mu^a$ and its inverse $e^\mu_a$
\begin{equation}
  \omega_{\mu}{}^a{}_b = e_\lambda^a e_b^\nu \Gamma^{\lambda}{}_{\mu\nu} -  e_b^\nu\partial_{\mu}e_\nu^a,
\end{equation} where $a$ and $b$ are indices in the defining representation of the local group $GL(4)$.
 In particular, \mbox{$GL(4) = \mathbb{R}_+ \times SL(4)$} and using \mbox{$SL(4) \simeq SO(3,3)$,} we can define spinorial representations for the diffeomorphisms group, eventually we will be able to define an action for $SO(3,3)$ spinors in four dimensions (check Appendix~\ref{sec:matter}).

\subsection*{Acknowledgments}
  We thank to J. Zanelli for fruitful discussions, and also to K. Peeters for helpful advises in the manipulation of the software \textsc{Cadabra}~\cite{Peeters:2007wn,peeters2007symbolic,Peeters2007550}, which was used extensively to achieve the results presented in this paper. Additionally, we thank the developers of the mathematical software \textsc{Sage}~\cite{sage}, used to achieve several manipulations. This work was partially supported by CONICYT (Chile) under grant No. 79140040.
%% \end{acknowledgments}

\appendix

\section{\label{sec:matter} Matter Fields}

The Dirac equation relies on the local $SO(3,1)$ Lorentz symmetry for everything. The aim of this section is to describe the inclusion of Dirac spinor without a local Lorentz symmetry in four dimensions. 

Dirac spinors are the fields that transform under representation of the local symmetry group that corresponds to the double cover of the original symmetry. We are interested in the representations of the diffeomorphisms group, that could be associated to the local symmetry generated by the semi-simple Lie group $SL(4,\mathbb{R})$ in four dimensions. This notion is unusual because typically one would think of $SO(3,1)$ as the local symmetry, and its double cover would generally be called the spin group $Spin(3,1)$. Instead, we realize that the local symmetry we have got, $SL(4,{\mathbb R})$ is equivalent to $SO(3,3)$ and the fundamental six dimensional representation of this group corresponds to the space of $4\times 4$ antisymmetric matrix representation of $SL(4,{\mathbb R})$. In terms of their invariant tensors, it is easy to see their correspondence. Consider a vector with components 
$v^A=\frac{1}{\sqrt{2}}(F^{01}-F^{23},F^{02}-F^{31},F^{03}-F^{12},F^{01}+F^{23},F^{02}+F^{31},F^{03}+F^{12})$, where $F^{ab}$ is an antisymmetric tensor representation of $SL(4)$. 
A diagonal metric $\eta_{AB} = \diag(-,-,-,+,+,+)$ allows us to compute inner products, and $\vec{v}\cdot\vec{v}$ is 
\begin{dmath}
  \vec{v}\cdot\vec{v} = 2\bigg(F^{01}F^{23}+F^{02}F^{31}+F^{03}F^{12}\bigg)
  =\frac{1}{4}\epsilon_{abcd}F^{ab}F^{cd}.
\end{dmath}
Thus, in order to consider $SL(4,{\mathbb R})$ double cover we will use the Clifford algebra ${\mathcal Cl}_{3,3}$ defined by 
\begin{equation}
  \comm{\Gamma_A}{\Gamma_B} = 2 \eta_{AB}
  \label{clifford}
\end{equation}
and redefine
\begin{dmath}
  \Gamma_A\rightarrow\frac{1}{\sqrt{2}}(\gamma_{01}-\gamma_{23},\gamma_{02}-\gamma_{31},\gamma_{03}-\gamma_{12},\gamma_{01}
  +\gamma_{23},\gamma_{02}+\gamma_{31},\gamma_{03}+\gamma_{12})
\end{dmath} 
to rewrite the Clifford algebra in Eq.~\eqref{clifford} as 
\begin{equation}
  \label{Clifford} 
  \big\{\gamma_{ab},\gamma_{cd}\big\}=2\epsilon_{abcd},
\end{equation}
where $\gamma_{ab}$ are antisymmetric in $a\leftrightarrow b$, $8\times 8$  complex matrices and $\eta_{AB}$ is basically $\epsilon_{abcd} $ in a different basis.

Thus, several topological scalars can be added to the Lagrangian density
\begin{equation}
  \label{SpinorLagrangian}
  \begin{split}
    \mathcal{L}_\Psi &= g_1 \bar\Psi\gamma_{ab}e^a_{\mu}e^b_{\nu}T^{\lambda,\mu\nu}\nabla_\lambda\Psi
    +g_2 \bar\Psi\gamma_{ab}e^a_{\mu}e^b_{\nu}\epsilon^{\lambda\kappa\mu\nu}A_\kappa\nabla_\lambda\Psi \\
    & \quad +g_3 \bar\Psi\gamma_{ab}e^a_{\mu}e^b_{\nu}\nabla_\lambda T^{\lambda,\mu\nu}\Psi
    +g_4 \bar\Psi\gamma_{ab}e^a_{\mu}e^b_{\nu}\epsilon^{\lambda\kappa\mu\nu}\nabla_\lambda A_\kappa\Psi.
  \end{split}
\end{equation}

%% \bibliographystyle{unsrt}
%% \bibliography{References.bib}

\end{document}